\documentclass[aps, twocolumn, superscriptaddress, showpacs,longbibliography]{revtex4-1}

\usepackage{amsmath,amsfonts,amssymb,amsthm}
\usepackage{hyperref}
\usepackage{bbm}
\usepackage{physics}
\usepackage{tabularx}
\usepackage{indentfirst}
\usepackage{caption}
\usepackage{subcaption}
\usepackage{tikz}
\usetikzlibrary{arrows.meta}

\makeatletter
\DeclareFontFamily{OMX}{MnSymbolE}{}
\DeclareSymbolFont{MnLargeSymbols}{OMX}{MnSymbolE}{m}{n}
\SetSymbolFont{MnLargeSymbols}{bold}{OMX}{MnSymbolE}{b}{n}
\DeclareFontShape{OMX}{MnSymbolE}{m}{n}{
	<-6>	MnSymbolE5
	<6-7>	MnSymbolE6
	<7-8>	MnSymbolE7
	<8-9>	MnSymbolE8
	<9-10>	MnSymbolE9
	<10-12>	MnSymbolE10
	<12->	MnSymbolE12
}{}
\DeclareFontShape{OMX}{MnSymbolE}{b}{n}{
	<-6>	MnSymbolE-Bold5
	<6-7>	MnSymbolE-Bold6
	<7-8>	MnSymbolE-Bold7
	<8-9>	MnSymbolE-Bold8
	<9-10>	MnSymbolE-Bold9
	<10-12>	MnSymbolE-Bold10
	<12->	MnSymbolE-Bold12
}{}

\let\llangle\@undefined
\let\rrangle\@undefined
\DeclareMathDelimiter{\llangle}{\mathopen}{MnLargeSymbols}{'164}{MnLargeSymbols}{'164}
\DeclareMathDelimiter{\rrangle}{\mathclose}{MnLargeSymbols}{'171}{MnLargeSymbols}{'171}
\makeatother

\makeatletter
\renewcommand\@makecaption[2]{%
 \par
 \vskip\abovecaptionskip
 \begingroup
 \small\rmfamily
 \begingroup
 \samepage
 \flushing
 \let\footnote\@footnotemark@gobble
 \@make@capt@title{#1}{#2}\par
 \endgroup
 \endgroup
 \vskip\belowcaptionskip
}
\makeatother

\newcommand{\Bra}[1]{\left\llangle #1 \right|}
\newcommand{\Ket}[1]{\left| #1 \right\rrangle}
\newcommand{\Braket}[2]{\left\llangle #1 \vphantom{#2} \right|\kern-0.6ex\left. #2 \vphantom{#1}\right\rrangle}
\newcommand{\EBra}[1]{\left[\left[ #1 \right|\right.}
\newcommand{\EKet}[1]{\left.\left| #1 \right]\right]}
\newcommand{\EBraket}[2]{\left[\left[ #1 \vphantom{#2} \right|\kern-0.6ex\left. #2 \vphantom{#1}\right]\right]}
\begin{document}

	\title{Einstein's quantum elevator: Hermitization of non-Hermitian Hamiltonians via a generalized vielbein formalism}

	\author{Chia-Yi Ju}
	\affiliation{Department of Physics, National Sun Yat-sen University, Kaohsiung 804, Taiwan}
	\author{Adam Miranowicz}
	\affiliation{Institute of Spintronics and Quantum Information, Faculty of Physics, Adam Mickiewicz University, 61-614 Pozna\'{n}, Poland}
	\affiliation{Theoretical Quantum Physics Laboratory, RIKEN Cluster for Pioneering Research, Wakoshi, Saitama 351-0198, Japan}
	\author{Fabrizio Minganti}
	\affiliation{Institute of Physics, Ecole Polytechnique F\'ed\'erale de Lausanne (EPFL), CH-1015 Lausanne, Switzerland}
	\affiliation{Theoretical Quantum Physics Laboratory, RIKEN Cluster for Pioneering Research, Wakoshi, Saitama 351-0198, Japan}
	\author{Chuan-Tsung Chan}
	\email{ctchan@go.thu.edu.tw}
	\affiliation{Department of Applied Physics, Tunghai University, Taichung 407, Taiwan}
	\author{Guang-Yin Chen}
	\email{gychen@phys.nchu.edu.tw}
	\affiliation{Department of Physics, National Chung Hsing University, Taichung 402, Taiwan}
	\author{Franco Nori}
	\affiliation{Theoretical Quantum Physics Laboratory, RIKEN Cluster for Pioneering Research, Wakoshi, Saitama 351-0198, Japan}
	\affiliation{RIKEN Center for Quantum Computing (RQC), Wakoshi, Saitama 351-0198, Japan}
	\affiliation{Department of Physics, University of Michigan, Ann Arbor, Michigan 48109-1040, USA}

	\begin{abstract}
		The formalism for non-Hermitian quantum systems sometimes blurs the underlying physics. We present a systematic study of the vielbein-like formalism which transforms the Hilbert space bundles of non-Hermitian systems into the conventional ones, rendering the induced Hamiltonian to be Hermitian. In other words, any non-Hermitian Hamiltonian can be ``transformed'' into a Hermitian one without altering the physics. Thus we show how to find a reference frame (corresponding to Einstein's quantum elevator) in which a non-Hermitian system, equipped with a non-trivial Hilbert space metric, reduces to a Hermitian system within the standard formalism of quantum mechanics.
	\end{abstract}

	\pacs{}
	\maketitle
	\newpage

	\section{Introduction}

		Since the discovery of $\cal{PT}$-symmetric quantum mechanics~\cite{Bender1998, Bender2002, Bender2004, Bender2007}, non-Hermitian quantum mechanics has become a very popular research field in quantum physics~\cite{Brody2016, ElGanainy2018, Bagarello2016, Ashida2020, Tzeng2021, Tu2022}. Even though the underlying mechanism of $\cal{PT}$-symmetric quantum mechanics was originally constructed from symmetry, many studies~\cite{Mostafazadeh2003, Mostafazadeh2004, Brody2013, Znojil2016, Ju2019} have pointed out that the Hilbert spaces of the quantum states require nontrivial metric operators in order to obtain self-consistent theories for non-Hermitian quantum systems. To be more precise, the quantum states live in fiber bundles, which hereinafter will be called Hilbert space bundles , where the fibers are Hilbert spaces and the base space is the time dimension in which the Hilbert spaces and states evolve.

		This dynamics can be better understood with the help of Einstein's elevator gedanken experiment. This gedanken experiment of a free-falling elevator laid the theoretical foundations for general relativity by showing the equivalence between inertial reference frames in a uniform gravitational field (curved spacetime) and accelerating reference frames, in which physical phenomena can be described within the gravitation-free (locally flat spacetime) special-relativity~\cite{Einstein2014, Chan2000}.

		We, therefore, ask an analogous question but concerning the ``equivalence relation'' between Hermitian and non-Hermitian formalisms of quantum mechanics: Does there exist a reference frame (a quantum version of Einstein's elevator) in which a non-Hermitian system, equipped with a non-trivial metric, reduces to a Hermitian system, within the standard formalism of quantum mechanics with a trivial metric operator (i.e., a metric that is an identity operator)? So the question is how to trivialize the metric of the Hilbert space bundle (i.e., to transform the metric into an identity operator) of a given non-Hermitian system. We constructively answer the question by applying a vielbein-like formalism~\cite{Nakahara2003} (see Appendix~\ref{VielbeinReview} for a brief review of the standard vielbein formalism in Riemannian geometry). This could be explained intuitively as breaking the metric into two pieces and spreading those into the vector space and its dual space so that the system seems trivial everywhere (see Fig.~\ref{FlattenFig}). Namely, the essence of the generalized vielbein formalism is simply linearly rearranging the vector and dual spaces so that the metric operator in the resulting space is an identity operator.

		\begin{figure*}[t!]
			\begin{minipage}{0.48\textwidth}
				{
					\centering
					\begin{tikzpicture}
						\def\i{250};
						\def\f{310};
						\def\r{2};
						\def\d{0.05};
						\def\u{6};
						\def\v{6};
						\foreach \m in {1, 2, ..., \u}{
							\draw[domain = 0: 1, variable = \x, smooth] plot ({-3 + \r * (cos(\x * (\f - \i) + \i) - cos(\i))}, {0.9 - 2 * (\m - 1) / (\u - 1) + \r * (sin(\x * (\f - \i) + \i) - sin(\i) + \d * sin(2 * \x * 360))});
							\draw[smooth] (1, {1 - 2 * (\m - 1) / (\u - 1)}) -- (3, {1 - 2 * (\m - 1) / (\u - 1)});
						}
						\foreach \n in {1, 2, ..., \v}{
							\draw ({-3 + \r * (cos(\i + (\f - \i) * (\n - 1) / (\v - 1)) - cos(\i))}, {0.9 + \r * (sin((\n - 1) / (\v - 1) * (\f - \i) + \i) - sin(\i) + \d * sin(2 * (\n - 1) / (\v - 1) * 360))}) -- ({-3 + \r * (cos(\i + (\f - \i) * (\n - 1) / (\v - 1)) - cos(\i))}, {-1.1 + \r * (sin((\n - 1) / (\v - 1) * (\f - \i) + \i) - sin(\i) + \d * sin(2 * (\n - 1) / (\v - 1) * 360))});
							\draw ({1 + 2 * (\n - 1) / (\v - 1)}, 1) -- ({1 + 2 * (\n - 1) / (\v - 1)}, -1);
						}
						\node at (-2, 1.4) {Original Basis};
						\node at (2, 1.4) {New Basis};
						\draw[<->, line width = 2] (-0.75, 0) arc (120: 60: 1.5) node[midway, above] {Vielbein};
						\node at (0, -0.1) {map};
					\end{tikzpicture}
				}
				\caption{An illustration of ``flattening'' the coordinates via the vielbein formalism. This procedure flattens the space (or curve in 1D). (Note that Regge calculus also flattens the curved manifold into flat space with deficit angles, which measure the local curvature~\cite{Misner2017}.)}
				\label{FlattenFig}
			\end{minipage} \quad%
			\begin{minipage}{0.48\textwidth}
				{
					\centering
					\begin{tikzpicture}
						\begin{scope}[shift={(-2.5, 0)}]
							\draw[line width = 1] (0, -1) -- (0, 0);
							\draw[fill = gray!20!white] (0, 0) ellipse (1.5 and 0.5);
							\draw[-latex, line width = 1] (0, 0) -- (0, 1);
							\draw[-{Latex[scale = 1.1]}, color = gray!50!black] (1, 0) -- (1, 0.8) node[above] {$\vec{v}$};
							\draw[fill = black] (1, 0) circle (0.05);
							\node at (0, 1.4) {Stationary Coordinate};
						\end{scope}
						\begin{scope}[shift={(2.5, 0)}]
							\draw[line width = 1] (0, -1) -- (0, 0);
							\draw[fill = gray!20!white] (0, 0) ellipse (1.5 and 0.5);
							\draw[-latex, line width = 1] (0, 0) -- (0, 1);
							\draw[->, line width = 0.5, domain = -225: 45, variable = \x] plot ({0.3 * cos(\x}, {0.7 + 0.15 * sin(\x)});
							\draw[-{Latex[scale = 1.1]}, color = gray!50!black, densely dashed] (1, 0) -- (1.8, 00) node[above] {$\vec{F}_\text{f}$};
							\draw[-{Latex[scale = 1.1]}, color = gray!50!black] (1, 0) -- (1, 0.6) node[above] {$\vec{v}'$};
							\draw[fill = black] (1, 0) circle (0.05);
							\node at (0.5, 1) {$\omega$};
							\node at (0, 1.4) {Rotating Coordinate};
						\end{scope}
						\draw[<->, line width = 1] (-0.8, 0) -- (0.8, 0);
						\node at (0, -0.7) {Coordinate Change};
					\end{tikzpicture}
				}
				\caption{Fictitious force induced from the coordinate change. This is a classical analog of different gauge choices inducing different Hamiltonians. As usual, there is no best gauge choice for all physical systems.}
				\label{GaugeFig}
			\end{minipage}
		\end{figure*}

		We can choose some bases~\cite{Nakahara2003}, by generalizing the vielbein technique, to simplify the calculations and obtain new insights into the systems under study. Indeed, the vielbein formalism is useful in many fields of physics, including general relativity~\cite{Wald1984, Misner2017}, supergravity~\cite{Nieuwenhuizen1981, West1990, Siegel1993}, superstring theories~\cite{Siegel1993a, Polacek2014, Ju2016, Linch2021}, etc. The main reason for applying the vielbein formalism is that it maps non-trivial space phenomena into a simpler space (and back).

		Since the metrics of the Hilbert space bundles of non-Hermitian quantum systems are not trivial, it is useful to study the vielbeins in these systems. Although some rudimentary ideas regarding the generalized vielbein formalism have been studied~\cite{Znojil2008, Mostafazadeh2010, Hamazaki2020, Ohlsson2021}, we here provide a clearer geometric understanding of this formalism. With the vielbein formalism, the time evolution of the transformed states is always governed by an induced Hermitian Hamiltonian.

		The formalism of non-Hermitian quantum mechanics, compared with that of standard quantum mechanics is quite complicated, as it requires one to calculate a non-trivial metric and its evolution for a given system. Otherwise, the omission of the metric can lead to apparent violations of fundamental principles in physics, including various no-go theorems in quantum information as explicitly explained in~\cite{Ju2019, Brody2016, Znojil2016}; indeed, the apparent violation of the non-signaling theorem~\cite{Lee2014}, the discrimination of non-orthogonal states~\cite{Pati2014}, and the increase in quantum entanglement by local operations~\cite{Chen2014} were reported.

		The actual usefulness of applying the proposed technique is to simplify calculations by reducing the formalism of non-Hermitian quantum mechanics, based on a non-trivial system-dependent metric, to the standard one of quantum mechanics with a trivial system-independent metric.

		In addition to how the formalism works, we also study the relations between different choices of vielbeins. This leads to a gauge transformation~\cite{Stefano2019, Garziano2020, Settineri2021, Savasta2021} which does not affect the physics. With different choices of gauges, the states evolve with different induced Hermitian Hamiltonians, which simplify calculations but do not alter the final physical results.

		A classical mechanics analogy of the gauge choice is a rotating or accelerating frame, which causes a fictitious force. The induced Hamiltonian plays a similar role to that of those fictitious forces in the time-dependent frame (see Fig.~\ref{GaugeFig}).

		In fact, the widely used Heisenberg and interaction pictures in Hermitian quantum mechanics are merely different choices of generalized vielbeins. After the construction of the generalized vielbein formalism in the Hilbert space bundles of quantum states and its gauge symmetry, we show some examples of the generalized vielbein formalism including how the Heisenberg and interaction pictures are related to the vielbeins.

	\section{From a metric to a vielbein.}

		Unlike Hermitian quantum systems, where the inner product between two states in Hilbert space is the familiar $\braket{\phi}{\psi}$, the Hilbert spaces of non-Hermitian quantum systems can have additional geometric structures so that the inner products in Hilbert space become $\Braket{\phi}{\psi} = \bra{\phi} G \ket{\psi}$, where $\Bra{\phi} = \bra{\phi} G$ is the corresponding dual state of $\ket{\phi}$ in the metricized Hilbert space with a metric $G$ (see Table~\ref{ComapringInnerProducts} for an explicit example). $G$ has to be Hermitian and positive-definite for a proper Hilbert space.

		In addition to the Hilbert space constraints mentioned above, this metric should also be constrained by the physics. The Hilbert space metric can be found by treating Schr\"{o}dinger's equation as a parallel transport~\cite{Ju2019}. It has been shown that if the Schr\"{o}dinger equation of a system is
		\begin{align}
			\partial_t \ket{\psi} = - i H \ket{\psi}, \label{OriginalSchroedingerEq}
		\end{align}
		where $H$ is its Hamiltonian, the compatibility of the metric $G$ with the Schr\"{o}dinger equation leads to
		\begin{align}
			\partial_t G + i H^\dagger G - i G H = 0. \label{MetricEq}
		\end{align}
		Although the solution for $G$ is not unique, they all differ by a gauge transformation, ${G \rightarrow G' = T^\dagger G T}$, where $T$ satisfies ${\partial_t T + i H T - i T H = 0}$.

		\begin{table}[h]
			\renewcommand*{\arraystretch}{1.6}
			\begin{tabular}{| >{\centering\arraybackslash}m{0.19\textwidth} | >{\centering\arraybackslash}m{0.25\textwidth} |}
				\hline
				Conventional & Metricized\\
				\hline
				\footnotesize $\braket{\phi}{\psi} = \begin{pmatrix}
					\phi_{1}^* & \phi_2^*
				\end{pmatrix}\begin{pmatrix}
					\psi_1\\ \psi_2
				\end{pmatrix}$ & \footnotesize \begin{tabular}{l}
					$\Braket{\phi}{\psi} = \bra{\phi} G \ket{\psi}$\\
					$= \begin{pmatrix}
						\phi_{1}^* & \phi_2^*
					\end{pmatrix}\begin{pmatrix}
						g_{11} & g_{12}\\
						g_{21} & g_{22}
					\end{pmatrix}\begin{pmatrix}
						\psi_1\\ \psi_2
					\end{pmatrix}$
				\end{tabular}\\
				\hline
			\end{tabular}
			\caption{Comparing the two inner products of two dimensional Hilbert spaces. In the conventional inner product, the dual state is just the Hermitian conjugate of the state; the dual state in the metricized space carries an additional metric operator $G$. Note that in the Hermitian case, the $G$ can always be chosen to be the identity which reduces back to the conventional space~\cite{Ju2019}.}
			\label{ComapringInnerProducts}
		\end{table}

		\begin{table*}[t]
			\renewcommand*{\arraystretch}{1.6}
			\begin{tabular}{| >{\centering\arraybackslash}m{0.18\textwidth} | >{\centering\arraybackslash}m{0.39\textwidth} | >{\centering\arraybackslash}m{0.39\textwidth} |}
				\hline
				& Original & Hermitized\\
				\hline
				State and dual state & $\Ket{\psi} = \ket{\psi}$\quad and\quad $\Bra{\psi} = \bra{\psi} G$ & $\EKet{\psi} = \mathcal{E} \ket{\psi}$\quad and\quad $\EBra{\psi} = \bra{\psi} \mathcal{E}^\dagger$\\
				\hline
				Inner product & $\Braket{\phi}{\psi} = \bra{\phi} G \ket{\psi}$ & $\EBraket{\phi}{\psi} = \bra{\phi} \mathcal{E}^\dagger \mathcal{E} \ket{\psi}$\\
				\hline
				Expectation values & $\left< \mathcal{O} \right> = \Bra{\psi} \mathcal{O} \Ket{\psi}$ & $\left< \mathcal{O} \right> = \EBra{\psi} \mathcal{O}_\flat \EKet{\psi} = \EBra{\psi} \mathcal{E} \mathcal{O} \mathcal{E}^{-1} \EKet{\psi}$\\
				\hline
				Governing Equations & \begin{tabular}{c}
						$\partial_t \Ket{\psi} = - i H \Ket{\psi}$,\quad\quad $\partial_t \Bra{\psi} = i \Bra{\psi} H$,\\
						$\partial_t G = i \left( G H - H^\dagger G \right)$
					\end{tabular} & \begin{tabular}{c}
						$\partial_t \EKet{\psi} = - i H_\flat \EKet{\psi}$,\quad\quad $\partial_t \EBra{\psi} = i \EBra{\psi} H_\flat$,\\
						$H_\flat = \mathcal{E} H \mathcal{E}^{-1} + i \left( \partial_t \mathcal{E} \right) \mathcal{E}^{-1}$
				\end{tabular}\\
				\hline
			\end{tabular}
			\caption{Relations between the original Hilbert space bundle and the Hermitized Hilbert space bundle. Since the metric $G$ is Hermitian and positive-definite, it can always be decomposed into $G = \mathcal{E}^\dagger \mathcal{E}$. }
			\label{Comparisons}
		\end{table*}

		Even though the metric carries the information of the Hilbert space bundle geometry, like the cases mentioned previously, it is not always desirable to keep the metric explicitly. A systematic way of ``removing'' the metric is to adopt the vielbein formalism, so that vectors are ``trivial'' everywhere.

		Since $G$ is Hermitian and positive-definite, it can be decomposed into
		\begin{align}
			G = \mathcal{E}^\dagger \mathcal{E}, \label{G=EE}
		\end{align}
		where the operator $\mathcal{E}$ is the generalized vielbein (which will be shortened to ``vielbein'' without risk of confusion).

		We can, therefore, use Eq.~\eqref{G=EE} to redefine the states in a locally Hermitian frame (in analogy to a locally flat frame). To be more specific, we define the state in the locally Hermitian frame to be
		\begin{align}
			\EKet{\psi} = \mathcal{E} \ket{\psi}. \label{NonCoodBasis}
		\end{align}

		It is obvious that the state evolution is no longer the same as Eq.~\eqref{OriginalSchroedingerEq}. A simple calculation shows that the time evolution of the new state is
		\begin{align}
			\partial_t \EKet{\psi} = - i H_\flat \EKet{\psi},
		\end{align}
		where
		\begin{align}
			H_\flat = \mathcal{E} H \mathcal{E}^{-1} + i \left( \partial_t \mathcal{E} \right) \mathcal{E}^{-1} \label{Hflat}
		\end{align}
		is the induced Hamiltonian. A quick calculation shows that Eq.~\eqref{Hflat} guarantees that $H_\flat = H_\flat^\dagger$ (see Appendix~\ref{HermiticityOfH} for a detailed derivation). This means that the induced Hamiltonian through the vielbein formalism, $H_\flat$, is always Hermitian.

		The induced Hamiltonian being Hermitian implies that the dual state of $\EKet{\psi}$ is the direct Hermitian conjugate of the state, i.e., $\EBra{\psi} = \left( \EKet{\psi} \right)^\dagger = \bra{\psi} \mathcal{E}^\dagger$. Hence, the inner product of the states in the locally Hermitian frame is reduced back to the usual inner product in the Hermitian system, while implicitly preserving the geometry of the Hilbert space bundle, i.e.,
		\begin{align}
			\EBraket{\psi_1}{\psi_2} = \bra{\psi_1} G \ket{\psi_2} = \Braket{\psi_1}{\psi_2}.
		\end{align}	

		Therefore, the vielbein $\mathcal{E}$ transforms any non-Hermitian Hamiltonians to Hermitian ones. This is true even at exceptional points (EPs)~\cite{Kato1976, Heiss2004, Oezdemir2019}.

	\section{Observables}

		The expectation value of an operator $\mathcal{O}$ is
		\begin{align}
			\begin{split}
				\left< \mathcal{O} \right> = & \Bra{\psi} \mathcal{O} \Ket{\psi} = \bra{\psi} G \mathcal{O} \ket{\psi}\\
				= & \bra{\psi} \mathcal{E}^\dagger \mathcal{E} \mathcal{O} \ket{\psi} = \EBra{\psi} \mathcal{E} \mathcal{O} \mathcal{E}^{-1} \EKet{\psi}.
			\end{split}
		\end{align}
		This means that the operator $\mathcal{O}$ in the locally Hermitian frame becomes $\mathcal{O}_\flat = \mathcal{E} \mathcal{O} \mathcal{E}^{-1}$.

		Using the fact that a self-adjoint operator $\mathcal{O}$ in the original space satisfies $\mathcal{O}^\dagger G = G \mathcal{O}$, together with Eq.~\eqref{G=EE} we find that
		\begin{align}
			\mathcal{O}_\flat^\dagger = & \left(\mathcal{E}^{-1}\right)^\dagger \mathcal{O}^\dagger \mathcal{E}^\dagger = \mathcal{E} \mathcal{O} \mathcal{E}^{-1} = \mathcal{O}_\flat,
		\end{align}
		i.e., the corresponding observable is also Hermitian in the locally Hermitian frame. Moreover, since $\mathcal{O}_\flat$ is merely a similarity transformation of $\mathcal{O}$, the eigenvalues of $\mathcal{O}_\flat$ are identical to those of $\mathcal{O}$. 

		Some comparisons between the original Hilbert space bundle and the Hermitized one are listed in Table~\ref{Comparisons}.

	\section{A Hidden Symmetry}

		The vielbein $\mathcal{E}$ is obtained from Eq.~\eqref{G=EE}, hence, by construction there are some inherited gauge symmetries from the metric $G$. Nevertheless, in parallel to the case in differential geometry, the vielbein introduced here also has more gauge freedoms than the metric. The ``gauge transformation'', a local transformation in $t$, for the vielbein is a unitary transformation, i.e., ${\mathcal{E} \rightarrow \mathcal{E}' = U \mathcal{E}}$, where $U$ is any unitary operator (including time-dependent ones). This additional gauge choice comes from the invariance of $G$; to be more specific, ${G' = \mathcal{E}^{\prime \dagger} \mathcal{E}' = \mathcal{E}^\dagger \mathcal{E} = G}$.

		Since $U$ can be any unitary operator, it can be generated by $\partial_t U = - i H_\text{L} U + i U H_\text{R}$, where $H_\text{L}$ and $H_\text{R}$ are Hermitian operators with ${U(t = 0)}$ being unitary. Then Eq.~\eqref{Hflat} for $\mathcal{E}'$ becomes
		\begin{align}
			\begin{split}
				H'_\flat = & ~ \mathcal{E}' H \mathcal{E}^{\prime-1} + i \left( \partial_t \mathcal{E}' \right) \mathcal{E}^{\prime-1}\\
				= & ~ H_\text{L} + U \left( H_\flat - H_\text{R}\right) U^{-1}.
			\end{split}
		\end{align}
		The detailed derivation can be found in Appendix~\ref{GaugeTransform}.

		Not only does this result show that the induced Hamiltonian depends on the gauge choice, but also it shows that $H_\flat$ can be chosen freely. To be more specific, given an $\mathcal{E}_1$ which induces a Hamiltonian $H_{1 \flat}$, we can make a gauge transformation, $U_{21}$, such that the states evolution is governed by $H_{2 \flat}$ for $U_{21}$ satisfying ${\partial_t U_{21} = - i H_{2 \flat} U_{21} + i U_{21} H_{1 \flat}}$.

		This means that even though $H_\flat$ governs the dynamics of the corresponding $\EKet{\psi}$, it is, in fact, telling us \emph{how the gauge choice evolves with time without altering the physics.}

		This gauge transformation might seem redundant at first; nevertheless, this has been often used in Hermitian quantum mechanics already. For example, the Heisenberg picture and interaction picture are special cases of the vielbein formalism with special choices of gauges [$H_\flat = 0$ and $H_\flat = H_\text{I}(t)$ respectively]. With this tool, we can ``remove'' the non-Hermiticity of the Hamiltonians.

	\section{Examples}

		To show how the vielbein formalism works, some examples, both in Hermitian and in non-Hermitian systems, are provided in the following.

		\subsection{Example 1: The Heisenberg picture and interaction pictures as gauge choices}

			The Heisenberg picture is historically the first picture of quantum mechanics extensively applied in many Hermitian studies. The main idea of the technique is to move all the time-dependence to the operator but leave the states time-independent. For the sake of clarity, we keep the time dependence of the states and operators explicit here.

			To achieve this, one first finds a unitary operator $\mathcal{U}_\text{H} (t)$, satisfying $\partial_t \mathcal{U}_\text{H} (t) = -i H (t) \mathcal{U}_\text{H} (t)$, and $\mathcal{U}_\text{H} (0) = \mathbbm{1}$. It is well-known that the states and the operators in the Heisenberg picture are defined to be ${\ket{\psi}_\text{H} = \ket{\psi (0)}}$ and ${\mathcal{O}_\text{H} (t) = \mathcal{U}_\text{H}^\dagger (t) \mathcal{O} (t) \mathcal{U}_\text{H} (t)}$, so that the operators carry all the time dependence while states have none, while leaving the physics unaltered, namely,
			\begin{align}
				\left< \mathcal{O}\right> \! (t) & = \bra{\psi (t)} \mathcal{O} (t) \ket{\psi (t)} = _\text{H}\hspace{-0.12cm}\bra{\psi} \mathcal{O}_\text{H} (t) \ket{\psi}_\text{H}.
			\end{align}

			To show that this is, in fact, a special case of the vielbein formalism, we let the vielbein be ${\mathcal{E} (t) = \mathcal{U}_\text{H}^{-1} (t) = \mathcal{U}_\text{H}^\dagger (t)}$. Therefore the states are
			\begin{align}
				\EKet{\psi(t)} = & ~ \mathcal{E}(t) \ket{\psi (t)} = \mathcal{U}_\text{H}^{-1} (t) \ket{\psi (t)} = \ket{\psi}_\text{H},
			\end{align}
			and the observables are
			\begin{align}
				\begin{split}
					\mathcal{O}_\flat (t) = & ~ \mathcal{E}(t) \mathcal{O} (t) \mathcal{E}^{-1} (t) = \mathcal{U}_\text{H}^{-1} (t) \mathcal{O} (t) \mathcal{U}_\text{H} (t) \\
					= & ~ \mathcal{U}_\text{H}^\dagger (t) \mathcal{O} (t) \mathcal{U}_\text{H} (t) = \mathcal{O}_\text{H}.
				\end{split}
			\end{align}
			The induced Hamiltonian is
			\begin{align}
				H_\flat (t) = & ~ \mathcal{E} (t) H (t) \mathcal{E}^{-1} (t) + i \left( \partial_t \mathcal{E} (t) \right) \mathcal{E}^{-1} (t) = 0.
			\end{align}
			Hence, the Heisenberg picture is the same as choosing the vielbein satisfying $\mathcal{E}(0) = \mathbbm{1}$, with the induced Hamiltonian $H_\flat = 0$.

			The interaction picture, on the other hand, is a different gauge choice, where $H_\flat = H_\text{I}(t)$. The detailed derivation can be found in Appendix~\ref{InteractionPicture}.

		\subsection{Example 2: A non-Hermitian case}

			Here, we demonstrate how the vielbein formalism works using the following Hamiltonian~\cite{Minganti2019}:
			\begin{align}
				H = \frac{\omega}{2} \sigma_x - i \frac{\gamma}{2} \sigma^+ \sigma^-,
			\end{align}
			where $\omega \neq 0$ and $\widetilde{\gamma} = \gamma / \omega$. The Hilbert space in this example is finite dimensional, but it can still be used in infinite-dimensional Hilbert space cases as well (see Appendix~\ref{InfiniteExmaple} for an infinite dimensional example).

			We split the discussion of this Hamiltonian into three cases [for $|\gamma| < 2 |\omega|$, $|\gamma| > 2 |\omega|$, and $\gamma = \pm 2 \omega$], because the Hamiltonian is non-diagonalizable at $\gamma = \pm 2 \omega$. In addition, we use three different starting points (the metric $G$, the vielbein $\mathcal{E}$ for $H_\flat = 0$, and the vielbein $\mathcal{E}$ for $H_\flat \neq 0$) in these three cases to show that they are mathematically equivalent.

			\subsubsection{Case $|\gamma| < 2 |\omega|$}

				We start with the metric method in this case. Solving Eq.~\eqref{MetricEq} together with $G$ being Hermitian and positive-definite, we find the metric
				\begin{align}
					\begin{split}
						G = & ~e^{\gamma t / 2} \begin{pmatrix}
							f^* & f\\
							\left(- i \frac{\gamma}{2 \omega} + \lambda_< \right) f^* & \left( - i \frac{\gamma}{2 \omega} - \lambda_< \right) f
						\end{pmatrix}\\
						& \times \begin{pmatrix}
							g_{11} & g_{12}\\
							g_{21} & g_{22}
						\end{pmatrix}
						\begin{pmatrix}
							f & \left(i \frac{\gamma}{2 \omega} + \lambda_< \right) f\\
							f^* & \left(i \frac{\gamma}{2 \omega} - \lambda_< \right) f^*
						\end{pmatrix},
					\end{split}
				\end{align}
				where $f = \exp\left( i \lambda_< \omega t / 2 \right)$, $\lambda_< = \left[ 1 - \gamma^2 / (2 \omega)^2 \right]^{1 / 2}$, and the $g_{ij}$'s are constants, such that $g_{11} > 0$, $g_{22} > 0$, $g_{12}^* =g_{21}$, and $|g_{12}|^2 < g_{11} g_{22}$.

				Using Eq.~\eqref{G=EE}, we find the corresponding $\mathcal{E}$ being
				\begin{align}
					\mathcal{E} = & ~ e^{\gamma t / 4} \begin{pmatrix}
						h_{11} & h_{12}\\
						h_{21} & h_{22}
					\end{pmatrix} \begin{pmatrix}
						f & \left(i \frac{\gamma}{2 \omega} + \lambda_< \right) f\\
						f^* & \left(i \frac{\gamma}{2 \omega} - \lambda_< \right) f^*
					\end{pmatrix},
				\end{align}
				where $g_{ij} = \sum_k h_{ik} h^*_{kj}$. Note that the $h_{ij}$'s can be time-dependent functions despite the $g_{ij}$'s being constants.

				Nevertheless, we first treat the $h_{ij}$'s as constants, using Eq.~\eqref{Hflat}, and find the induced Hamiltonian $H_\flat = 0$, since $\partial_t \mathcal{E} = i \mathcal{E} H$.

				We next make a gauge transformation to $\mathcal{E}' = U \mathcal{E}$, where
				\begin{align}
					U = \exp \left( - i \dfrac{\omega t}{2} \sigma_x \right) = \begin{pmatrix}
						\cos\dfrac{\omega t}{2} & - i \sin\dfrac{\omega t}{2}\\
						- i \sin\dfrac{\omega t}{2} & \cos\dfrac{\omega t}{2}
					\end{pmatrix}.
				\end{align}
				A direct calculation shows that the induced Hamiltonian of $\mathcal{E}'$ is then $H'_\flat = \omega \sigma_x / 2$. So that if $\mathcal{E}'$ is chosen to be the vielbein, the state evolution is governed by ${\partial_t \EKet{\psi} = H'_\flat \EKet{\psi}}$.

			\subsubsection{2. Case $|\gamma| > 2 |\omega|$}

				To find the corresponding metric $G$, we could calculate the metric using Eq.~\eqref{MetricEq} as we did in the previous case. Nevertheless, this time we start with Eq.~\eqref{Hflat}, while letting $H_\flat = 0$ so the general solution of the vielbein becomes
				\begin{align}
					\mathcal{E} = e^{\gamma t / 4} \begin{pmatrix}
						h_{11} & h_{12}\\
						h_{21} & h_{22}
					\end{pmatrix} \begin{pmatrix}
						f^+ & i \left(\frac{\gamma}{2 \omega} - \lambda_> \right) f^+\\
						f^- & i \left(\frac{\gamma}{2 \omega} + \lambda_> \right) f^-
					\end{pmatrix},
				\end{align}
				where $f^\pm = \exp\left(\pm \lambda_> \omega t / 2\right)$, $\lambda_> = \left[\gamma^2 / \left(4 \omega^2 \right) - 1\right]^{1 / 2}$, and the matrix of $h_{ij}$'s is constant with a non-vanishing determinant.

				We can then use Eq.~\eqref{G=EE} to find that
				\begin{align}
					\begin{split}
						G = & ~ e^{\gamma t / 2} \begin{pmatrix}
							f^+ & f^-\\
							- i \left(\frac{\gamma}{2 \omega} - \lambda_> \right) f^+ & - i \left(\frac{\gamma}{2 \omega} + \lambda_> \right) f^-
						\end{pmatrix}\\
						& \times \begin{pmatrix}
							g_{11} & g_{12}\\
							g_{21} & g_{22}
						\end{pmatrix}
						 \begin{pmatrix}
							f^+ & i \left(\frac{\gamma}{2 \omega} - \lambda_> \right) f^+\\
							f^- & i \left(\frac{\gamma}{2 \omega} + \lambda_> \right) f^-
						\end{pmatrix},
					\end{split}
				\end{align}
				where $g_{ij} = \sum_k h_{ik} h^*_{kj}$. Now the $g_{ij}$'s are constants, such that $g_{11} > 0$, $g_{22} > 0$, $g_{12}^* = g_{21}$, and $|g_{12}|^2 < g_{11} g_{22}$, which indeed make the metric Hermitian and positive-definite.

				We can, again, apply a gauge transformation from $\mathcal{E}$ to $\mathcal{E}' = U \mathcal{E}$, with $U = \exp \left[ (- i / 2) \omega t \sigma_x \right]$. Direct calculation shows that the induced Hamiltonian also becomes ${H'_\flat = \omega \sigma_x / 2}$.

			\subsubsection{Exceptional point at $\gamma = \pm 2 \omega$}

				In this case, the Hamiltonian can be written as
				\begin{align}
					H_\text{EP} = \frac{\omega}{2} \begin{pmatrix}
						\mp 2 i & 1\\
						1 & 0
					\end{pmatrix},
				\end{align}
				which is non-diagonalizable and corresponds to an EP~\cite{Oezdemir2019}. Although we can still find its corresponding metric by solving Eq.~\eqref{MetricEq} directly, we start with Eq.~\eqref{Hflat} while setting $H_\flat = \omega \sigma_x / 2$. Thus, using Eq.~\eqref{Hflat}, we find
				\begin{align}
					\mathcal{E} = e^{\frac{\left(\pm \mathbbm{1}- i \sigma_x \right) \omega t}{2}}\begin{pmatrix}
						h_{11} & h_{12}\\
						h_{21} & h_{22}
					\end{pmatrix} \begin{pmatrix}
						1 & \pm i \\
						2 \omega t & i ( \pm 2 \omega t - 4 )
					\end{pmatrix},
				\end{align}
				where the constant matrix made of $h_{ij}$'s has a non-vanishing determinant.

				Using Eq.~\eqref{G=EE}, we, again, find the corresponding metric
				\begin{align}
					\begin{split}
						G = & ~ e^{\pm \omega t} \begin{pmatrix}
							1 & 2 \omega t\\
							\mp i & - i ( \pm 2 \omega t - 4 )
						\end{pmatrix}
						\begin{pmatrix}
							g_{11} & g_{12}\\
							g_{21} & g_{22}
						\end{pmatrix}\\
						& \times \begin{pmatrix}
							1 & \pm i \\
							2 \omega t & i ( \pm 2 \omega t - 4 )
						\end{pmatrix},
					\end{split}
				\end{align}
				where $g_{ij} = \sum_k h_{ik} h^*_{kj}$. Clearly we can always use ${U = \exp \left( i \omega t \sigma_x / 2 \right)}$ to transform the induced Hamiltonian into $H'_\flat = 0$.

	\section{Conclusion}

		We refer to the analogy between non-Hermitian quantum mechanics and general relativity having in mind both (i) the physical analogy between Hermitization of non-Hermitian Hamiltonians and Einstein's quantum elevator, and (ii) the mathematical analogy concerning the usage of the vielbein formalism. Indeed, this formalism is a powerful approach to general relativity. However, of course, its usefulness is not limited to this theory. It can be directly applied, e.g., to supergravity and superstring theories. Moreover, as shown here, it can also be useful to study the dynamics of non-Hermitian quantum mechanics.

		Compared with the standard coordinate-based approach to general relativity, the vielbein formalism enables finding an optimal vielbein basis to simplify the description of the spacetime and to reveal its specific physical aspects. Analogously, the vielbein-like formalism applied here simplifies the analysis of non-Hermitian quantum systems (with highly nontrivial metric operators in their Hilbert spaces) by mapping the problem to locally Hermitian ones with a standard (i.e., trivial) metric. After such a simplification, important aspects of the dynamics of non-Hermitian systems can be clearly and intuitively revealed.

		To be more specific, the non-trivial Hilbert space bundle metrics in non-Hermitian quantum systems sometimes complicate the physical description of these systems. Following the geometrical meaning of Schr\"{o}dinger's equation, we find that non-Hermitian Hamiltonians can be transformed into Hermitian ones via the vielbein formalism, shedding a new light on the physics of non-Hermitian systems. We also present a systematic study on an additional gauge symmetry which originates from the freedom of choosing vielbein frames, where the quantum states evolution is described by different Hamiltonians. Furthermore, the vielbein formalism is \emph{not} restricted to non-Hermitian quantum systems. The gauge freedoms in the vielbein formalism in Hermitian systems also grants us the freedom to choose frames, the Heisenberg and interaction pictures, for example, that are easier to work with. 

	\begin{acknowledgments}
		\emph{Acknowledgment.}--- C.-Y.J. is partially supported by the Ministry of Science and Technology (MOST) through Grant No. MOST 111-2112-M-110-007-MY2. G.Y.C. is partially supported by the National Center for Theoretical Sciences and the MOST through Grant No. MOST 110-2123-M-006-001 and No. MOST 110-2112-M-005-002. C.T.C. is partially supported by the MOST through Grant No. MOST 109-2112-M-029-006- and 110-2112-M-029-006-. A.M. is supported by the Polish National Science Centre (NCN) under the Maestro Grant No. DEC-2019/34/A/ST2/00081. F.N. is supported in part by: Nippon Telegraph and Telephone Corporation (NTT) Research, the Japan Science and Technology Agency (JST) [via the Quantum Leap Flagship Program (Q-LEAP) and the Moonshot R\&D Grant Number JPMJMS2061], the Japan Society for the Promotion of Science (JSPS) [via the Grants-in-Aid for Scientific Research (KAKENHI) Grant No. JP20H00134], the Army Research Office (ARO) (Grant No. W911NF-18-1-0358), the Asian Office of Aerospace Research and Development (AOARD) (via Grant No. FA2386-20-1-4069), and the Foundational Questions Institute Fund (FQXi) via Grant No. FQXi-IAF19-06.
	\end{acknowledgments}

	\begin{appendix}
		\section{A brief review of the standard vielbein formalism \label{VielbeinReview}}

			It is well-known that in a Riemannian geometry, the inner product between two vectors $A$ and $B$ is defined as
			\begin{align}
				\left< A, B \right> = A^\mu g_{\mu \nu} B^\nu, \label{StandardInnerProduct}
			\end{align}
			where $g_{\mu \nu} = g_{\nu \mu}$ is the metric tensor (component). Note that the Einstein summation rule is applied here. In general, the metric tensor can be very complicated and becomes an obstacle to the understanding of the geometry/physics. Nevertheless, we can always make a local coordinate transformation that renders the inner product formally simpler.

			The standard vielbein formalism in differential geometry~\cite{Nakahara2003} is a technique of finding such a transformation by breaking the metric tensor into the product of two vielbeins, namely,
			\begin{align}
				g_{\mu \nu} = e_\mu^{~a} \delta_{a b} e_\nu^{~b},
			\end{align}
			where $\delta_{a b}$ is the Kronecker delta and $e_\mu^a$ and $e_\nu^b$ are the vielbeins. Then we define the new vector components as
			\begin{align}
				\widetilde{A}^a = e_\mu^{~a} A^\mu \quad \text{and} \quad \widetilde{B}^b = e_\nu^{~b} B^\nu, \label{StandardNonCoordBasis}
			\end{align}
			so that
			\begin{align}
				\left< A, B \right> = \widetilde{A}^a \delta_{a b} \widetilde{B}^b,
			\end{align}
			which formally looks like the inner product in a flat space. For standard Riemannian geometry, applying the vielbein formalism does not alter the outcome of the geometric/physics theory, but makes the theory conceptually clearer. Nevertheless, there are some cases where the physics can only be described using the vielbein formalism but not the metric one, for example, when fermions are present in a gravitational theory.

			To show that the vielbein formalism implies a gauge symmetry (redundant degrees of freedom), we rewrite the metric and the vielbeins in terms of matrices
			\begin{align}
				g = \begin{pmatrix}
					g_{11} & g_{12} & \cdots\\
					g_{21} & g_{22} & \cdots\\
					\vdots & \vdots & \ddots
				\end{pmatrix} \quad \text{and} \quad e = \begin{pmatrix}
					e_1^{~1} & e_1^{~2} & \cdots\\
					e_2^{~1} & e_2^{~2} & \cdots\\
					\vdots & \vdots & \ddots
				\end{pmatrix},
			\end{align}
			which implies
			\begin{align}
				g = e^\top e. \label{g=eTe}
			\end{align}
			Since $g = g^\top$ is a symmetric real matrix, the degrees of freedom of $g$ is $n (n + 1) / 2$, where $n$ is the dimension of the manifold. Nevertheless, the vielbein $e$ does not have such a restriction and, therefore, is an element of GL($n$) which has $n^2$ degrees of freedom.

			It might seem that the degrees of freedom in the vielbein formalism are $n (n - 1) / 2$ degrees too many, yet the $e$ in Eq.~\eqref{g=eTe} is not uniquely defined. That is, if there is a $\Lambda$ such that
			\begin{align}
				\Lambda^\top \Lambda = \mathbbm{1},
			\end{align}
			i.e., $\Lambda \in$ SO($n$), then $e' = \Lambda e$ can also be a vielbein, namely
			\begin{align}
				e^{\prime \top} e = e^\top \Lambda^\top \Lambda e = e^\top e = g.
			\end{align}
			Therefore, there are some redundant SO($n$) degrees of freedom [exactly $n (n - 1) / 2$ degrees of freedom] in the vielbein formalism. Therefore, the vielbein formalism has an SO($n$) gauge symmetry.

			To make contact with the vielbein formalism provided in the main text, we further rewrite $A = (A^1 ~ A^2 ~ \cdots)^\top$ and $B = (B^1 ~ B^2 ~ \cdots)^\top$, and the inner product in Eq.~\eqref{StandardInnerProduct} becomes
			\begin{align}
				\left< A, B \right> = A^\top g B,
			\end{align}
			which is generalized to $\Braket{A}{B} = \bra{A} G \ket{B}$ by replacing $^\top$ with $^\dagger$ in the quantum mechanics case. Since $G^\dagger = G$, the analog of Eq.~\eqref{g=eTe} becomes Eq.~\eqref{G=EE}, and the transformations in Eq.~\eqref{StandardNonCoordBasis}, i.e.,
			\begin{align}
				\widetilde{A} = e A \quad \text{and} \quad \widetilde{B} = e B,
			\end{align}
			become Eq.~\eqref{NonCoodBasis}.

		\section{The Hermiticity of the induced Hamiltonian \label{HermiticityOfH}}

			The goal of this appendix is to show that the induced Hamiltonian in the vielbein formalism is always Hermitian.

			To show this, we use the fact that the time evolution equation of the metric $G$ is
			\begin{align}
				& \partial_t G = i \left( G H - H^\dagger G \right) \label{MetricEq}\\
				\Rightarrow & \partial_t \left(\mathcal{E}^\dagger \mathcal{E}\right) = i \left( \mathcal{E}^\dagger \mathcal{E} H - H^\dagger \mathcal{E}^\dagger \mathcal{E} \right).
			\end{align}

			A direct calculation shows that
			\begin{align}
				\begin{split}
					H_\flat^\dagger = & \left( \mathcal{E}^\dagger \right)^{-1} H^\dagger \mathcal{E}^\dagger - i \left( \mathcal{E}^\dagger \right)^{-1} \left( \partial_t \mathcal{E}^\dagger \right)\\
					= & \left( \mathcal{E}^\dagger \right)^{-1} H^\dagger \mathcal{E}^\dagger \left( \mathcal{E} \mathcal{E}^{-1} \right) - i \left( \mathcal{E}^\dagger \right)^{-1} \left( \partial_t \mathcal{E}^\dagger \right) \left( \mathcal{E} \mathcal{E}^{-1} \right)\\
					= & \left( \mathcal{E}^\dagger \right)^{-1} H^\dagger G \mathcal{E}^{-1} - i \left( \mathcal{E}^\dagger \right)^{-1} \left( \partial_t \mathcal{E}^\dagger \mathcal{E} \right) \mathcal{E}^{-1}\\
					& + i \left( \mathcal{E}^\dagger \right)^{-1} \mathcal{E}^\dagger \left( \partial_t \mathcal{E} \right) \mathcal{E}^{-1}\\
					= & \left( \mathcal{E}^\dagger \right)^{-1} H^\dagger G \mathcal{E}^{-1} - i \left( \mathcal{E}^\dagger \right)^{-1} \left( \partial_t G \right) \mathcal{E}^{-1} + i \left( \partial_t \mathcal{E} \right) \mathcal{E}^{-1}\\
					= & \left( \mathcal{E}^\dagger \right)^{-1} G H \mathcal{E}^{-1} + i \left( \partial_t \mathcal{E} \right) \mathcal{E}^{-1}\\
					= & ~ \mathcal{E} H \mathcal{E}^{-1} + i \left( \partial_t \mathcal{E} \right) \mathcal{E}^{-1} = H_\flat,
				\end{split}\label{HHermiticity}
			\end{align}
			where Eqs.~\eqref{G=EE} and \eqref{MetricEq} are applied in the derivation.

			Note that the gauge choice of vielbein $\mathcal{E}$ has not been specified. That is to say, any vielbein gauge choice renders $H_\flat = H_\flat^\dagger$. Hence the induced Hamiltonian via the vielbein formalism is always Hermitian.

		\section{The ``gauge transformation'' on $H_\flat$ \label{GaugeTransform}}

			This appendix focuses on the detailed proof of Eq.~(10) in the main text and its implication.

			It is known that the decomposition of Eq.~\eqref{G=EE} is far from unique. If $\mathcal{E}$ satisfies Eq.~\eqref{G=EE}, we can always find another vielbein $\mathcal{E}' = U \mathcal{E}$ that also satisfies $G = \mathcal{E}^{\prime \dagger} \mathcal{E}'$. By construction, we find
			\begin{align}
				& \mathcal{E}^\dagger \mathcal{E} = G = \mathcal{E}^{\prime \dagger} E' = \mathcal{E} U^\dagger U E\\
				& \Rightarrow U^\dagger U = \mathbbm{1}, \label{UnitaryTransformation}
			\end{align}
			which shows that $U$ can be any unitary operator.

			To show that the unitary group is \emph{the} gauge group or the redundant freedom for the system, we count the degrees of freedom in the metric $G$, the vielbein $E$, and the gauge transformation $U$ for a Hilbert space with dimension $n$. The metric $G$ is a complex matrix satisfying $G = G^\dagger$, which gives $n + 2 n (n - 1) / 2 = n^2$ degrees of freedom, where the first $n$ comes from the diagonal elements being real and the $2 n (n - 1) / 2$ comes from the upper off-diagonal elements because there are $n (n - 1) / 2$ complex elements. For the vielbein $E$, it is an element of GL($n, \mathbb{C}$), which has $2 n^2$ degrees of freedom, which is $n^2$ too many compared with the metric $G$. Nevertheless, the unitary group U($n$) has $n^2$ degrees of freedom, which is exactly the redundant degrees of freedom in the vielbein $E$. Together with Eq.~\eqref{UnitaryTransformation}, the redundant freedom in the vielbein is solely described by a unitary group.

			Using the fact that $U$ is unitary, the time derivative of $U$ can always be written as
			\begin{align}
				\partial_t U = - i H_\text{L} U + i U H_\text{R},
			\end{align}
			where $H_\text{L}$ and $H_\text{R}$ are Hermitian operators with ${U(t = 0)}$ being unitary. Using Eq.~\eqref{Hflat} for $\mathcal{E}'$, we find
			\begin{align}
				\begin{split}
					H'_\flat = & ~ E' H \mathcal{E}^{\prime-1} + i \left( \partial_t E' \right) \mathcal{E}^{\prime-1}\\
					= & ~ U \mathcal{E} H \mathcal{E}^{-1} U^{-1} + i \left( \partial_t U \mathcal{E} \right) \mathcal{E}^{-1} U^{-1}\\
					= & ~ U \mathcal{E} H \mathcal{E}^{-1} U^{-1} + H_\text{L} U \mathcal{E} \mathcal{E}^{-1} U^{-1}\\
					& ~ - U H_\text{R} \mathcal{E} \mathcal{E}^{-1} U^{-1}+ i U \left( \partial_t \mathcal{E} \right) \mathcal{E}^{-1} U^{-1}\\
					= & ~ H_\text{L} + U \left( H_\flat - H_\text{R}\right) U^{-1},
				\end{split}
			\end{align}
			which proves Eq.~(10) in the main text.

			This result shows that when we have a vielbein, say $\mathcal{E}_1$, that induces the Hamiltonian $H_{1 \flat}$, we can apply a gauge transform on $\mathcal{E}_1$ to $\mathcal{E}_2 = U_{21} \mathcal{E}_1$, such that the new induced Hamiltonian $H_{2 \flat}$ can be any given Hermitian operator. To achieve this, $U_{21}$ needs to be unitary at some time $t$ and its time derivative must satisfy
			\begin{align}
				\partial_t U_{21} = - i H_{2 \flat} U_{21} + i U_{21} H_{1 \flat}.
			\end{align}

			Hence, we can always choose a frame that is convenient to work with.

		\section{The interaction picture as a gauge choice \label{InteractionPicture}}

			Besides the Heisenberg picture shown in the main text, another standard picture in Hermitian quantum mechanics is the interaction picture, which is particularly useful in perturbation methods. We keep the time-dependence explicit in this appendix to avoid possible confusion.

			The main idea of the interaction picture is to split the Hamiltonian into a ``system'' part and an ``interaction'' part, namely,
			\begin{align}
				H (t) = H_\text{s} (t) + H_\text{int} (t),
			\end{align}
			where $H_\text{s} (t)$ and $H_\text{int} (t)$ are Hermitian.

			One then finds a $\mathcal{U}_\text{I} (t)$ such that
			\begin{align}
				\partial_t \mathcal{U}_\text{I} (t) = -i H_\text{s} (t) \mathcal{U}_\text{I} (t)
			\end{align}
			with $\mathcal{U}_\text{I} (0) = \mathbbm{1}$. It is obvious that the $\mathcal{U}_\text{I} (t)$ satisfying these two conditions is a unitary operator. The states and the operators in the interaction picture are defined to be
			\begin{align}
				& \ket{\psi (t)}_\text{I} = \mathcal{U}_\text{I}^{-1} (t) \ket{\psi (t)},\\
				& \mathcal{O}_\text{I} (t) = \mathcal{U}_\text{I}^\dagger (t) \mathcal{O} (t) \mathcal{U}_\text{I} (t).
			\end{align}
			It can be shown that the time evolution of the states in the interaction picture is
			\begin{align}
				i \partial_t \ket{\psi (t)}_\text{I} = H_\text{I}(t) \ket{\psi (t)}_\text{I},
			\end{align}
			where
			\begin{align}
				H_\text{I} (t) = \mathcal{U}_\text{I}^{-1} (t) H_\text{int} (t) \mathcal{U}_\text{I} (t).
			\end{align}

			Back to the vielbein formalism, we can choose the vielbein to be
			\begin{align}
				\mathcal{E} (t) = \mathcal{U}_\text{I}^{-1} (t) = \mathcal{U}_\text{I}^\dagger (t),
			\end{align}
			so that the states and the operators become
			\begin{align}
				\EKet{\psi(t)} = & ~ \mathcal{E}(t) \ket{\psi (t)} = \mathcal{U}_\text{I}^{-1} (t) \ket{\psi (t)} = \ket{\psi}_\text{I},\\
				\begin{split}
					\mathcal{O}_\flat (t) = & ~ \mathcal{E}(t) \mathcal{O} (t) \mathcal{E}^{-1} (t) = \mathcal{U}_\text{I}^{-1} (t) \mathcal{O} (t) \mathcal{U}_\text{I} (t)\\
					= & ~ \mathcal{U}_\text{I}^\dagger (t) \mathcal{O} (t) \mathcal{U}_\text{I} (t) = \mathcal{O}_\text{I}.
				\end{split}
			\end{align}
			The induced Hamiltonian in this case is
			\begin{align}
			 	\begin{split}
				 	H_\flat (t) & = \mathcal{E} (t) H (t) \mathcal{E}^{-1} (t) + i \left[ \partial_t \mathcal{E} (t) \right] \mathcal{E}^{-1} (t)\\
					& = \mathcal{U}_\text{I}^{-1} (t) H (t) \mathcal{U}_\text{I} (t) - i \mathcal{U}_\text{I}^{-1} (t) \partial_t \mathcal{U}_\text{I} (t)\\
					& = \mathcal{U}_\text{I}^{-1} (t) H (t) \mathcal{U}_\text{I} (t) - \mathcal{U}_\text{I}^{-1} (t) H_\text{s} (t) \mathcal{U}_\text{I} (t)\\
					& = \mathcal{U}_\text{I}^{-1} (t) H_\text{int} (t) \mathcal{U}_\text{I} (t)\\
					& = H_\text{I} (t).
				\end{split}
			\end{align}

			Therefore, the interaction picture in Hermitian quantum mechanics is also a special choice of the vielbein.

		\section{An infinite dimension non-Hermitian system example \label{InfiniteExmaple}}

			In this appendix, we demonstrate that this vielbein formalism also works for an infinite dimensional Hilbert space bundle with the Hamiltonian
			\begin{align}
				\begin{split}
					H = & - i \frac{\gamma_a}{2} a^\dagger a - i \frac{\gamma_b}{2} b^\dagger b + g \left( a^\dagger b + b^\dagger a\right)\\
					= & \begin{pmatrix}
							a^\dagger & b^\dagger
						\end{pmatrix} \begin{pmatrix}
							- i \dfrac{\gamma_a}{2} & g\\
							g & - i \dfrac{\gamma_a}{2} 
						\end{pmatrix} \begin{pmatrix}
							a\\
							b
						\end{pmatrix},
				\end{split}
			\end{align}
			where $a$ and $b$ ($a^\dagger$ and $b^\dagger$) are the bosonic annihilation (creation) operators. For later convenience, we define the vacuum state $\ket{0}$ such that
			\begin{align}
				a \ket{0} = \ket{0} \cdot 0 ~ = 0,\\
				b \ket{0} = \ket{0} \cdot 0 ~ = 0.
			\end{align}

			\subsection{Case without exceptional point}

				When $\left|\gamma_a - \gamma_b\right| \neq 4 |g|$, the Hamiltonian can be rewritten as
				\begin{align}
					\begin{split}
						H = & ~ h_+ c_+^\text{c} c_+^\text{a} + h_- c_-^\text{c} c_-^\text{a}\\
						= & \begin{pmatrix}
							c_+^\text{c} & c_-^\text{c}
						\end{pmatrix} \begin{pmatrix}
							h_+ & 0\\
							0 & h_-
						\end{pmatrix} \begin{pmatrix}
							c_+^\text{a}\\
							c_-^\text{a}
						\end{pmatrix},
					\end{split}
				\end{align}
				where
				\begin{align}
					\begin{split}
						h_\pm = & ~ - i \dfrac{\gamma_a + \gamma_b}{4} \pm \dfrac{\zeta}{4},\\
						c_\pm^\text{c} = & ~ \left[ a^\dagger \mp \dfrac{\zeta \mp i \left(\gamma_a - \gamma_b\right)}{4 g} b^\dagger \right],\\
						c_\pm^\text{a} = & ~ \frac{1}{\zeta} \left[ \dfrac{\zeta \pm i \left(\gamma_a - \gamma_b \right)}{2} a \mp 2 g b \right],\\
						\zeta^2 = & ~ 16 g^2 - \left( \gamma_a - \gamma_b \right)^2 .
					\end{split}
				\end{align}
				The commutation relations between $H$, $c^\text{c}$, and $c^\text{a}$ are
				\begin{align}
					\begin{split}
						& \left[ c_\pm^\text{a}, c_\pm^\text{c} \right] = 1,\\
						& \left[ c_\pm^\text{a}, c_\mp^\text{c} \right] = 0,\\
						& \left[ H, c_\pm^\text{c} \right] = h_\pm c_\pm^\text{c},\\
						& \left[ H, c_\pm^\text{a} \right] = - h_\pm c_\pm^\text{a}.
					\end{split}
				\end{align}

				Then by solving Eq.~\eqref{MetricEq}, together with $G = G^\dagger$ and positive-definiteness, we find
				\begin{align}
					\begin{split}
						G = & \sum_{\substack{n_+ = 0\\ n_- = 0 }}^\infty \frac{g_{n_+ n_-}}{\left(n_+ !\right)^2 \left(n_- !\right)^2} \exp \left[ -2 t \left( n_+ \Im h_+ + n_- \Im h_- \right) \right]\\
						& \cdot \left( c_-^{\text{a} \dagger} \right)^{n_-} \left( c_+^{\text{a} \dagger} \right)^{n_+} \ket{0} \! \bra{0} \left( c_+^\text{a} \right)^{n_+} \left( c_-^\text{a} \right)^{n_-},
					\end{split}
				\end{align}
				where $g_{ij} > 0$ are constants.

				To find the corresponding vielbeins, we only need to find one $\mathcal{E}$ that satisfies Eq.~\eqref{G=EE}; then, the general solution can be found by a simple gauge transformation. A vielbein that satisfies Eq.~\eqref{G=EE} is
				\begin{align}
					\begin{split}
						\mathcal{E} = & \sum_{\substack{n_+ = 0\\ n_- = 0 }}^\infty \frac{h_{n_+ n_-}}{\left(n_+ !\right)^{3 / 2} \left(n_- !\right)^{3 / 2}} \exp \left[ i t \left( n_+ h_+ + n_- h_- \right) \right]\\
						& \cdot \left( a^\dagger \right)^{n_-} \left( b^\dagger\right)^{n_+} \ket{0}\!\bra{0} \left( c_+^\text{a} \right)^{n_+} \left( c_-^\text{a} \right)^{n_-},
					\end{split}
				\end{align}
				where $h_{n_+ n_-}$'s are non-zero constants and ${g_{n_+ n_-} = \left| h_{n_+ n_-} \right|^2 > 0}$. A direct calculation shows that the induced Hamiltonian is $H_\flat = 0$.

				We can make a gauge transformation on $\mathcal{E}$ to $\mathcal{E}'$ by
				\begin{align}
					\mathcal{E}' = U \mathcal{E}, \label{InfiniteE}
				\end{align}
				where
				\begin{align}
					\begin{split}
						U = & \sum_{\substack{m = 0\\ n = 0 }}^\infty \frac{\exp \left[itg(m - n) \right]}{\sqrt{2}^{m + n} \left(m !\right) \left(n !\right)}\\
						& \cdot \left( a^\dagger + b^\dagger \right)^{m} \left( a^\dagger - b^\dagger \right)^{n} \ket{0}\!\bra{0} a^{m} b^{n},
					\end{split}
				\end{align}
				Then the induced Hermitian Hamiltonian becomes
				\begin{align}
					H'_\flat = g (a^\dagger b + a b^\dagger),
				\end{align}
				if the vielbein is chosen to be $\mathcal{E}'$ in Eq.~\eqref{InfiniteE}, namely,
				\begin{align}
					\begin{split}
						\mathcal{E}' = & \sum_{\substack{n_+ = 0\\ n_- = 0 }}^\infty \frac{h_{n_+ n_-} \exp \left\lbrace i t \left[ n_+ \left(h_+ + g \right) + n_- \left(h_- - g \right) \right] \right\rbrace}{\sqrt{2}^{n_+ + n_-} \left(n_+ !\right)^{3 / 2} \left(n_- !\right)^{3 / 2}}\\
						& \cdot \left( a^\dagger + b^\dagger \right)^{n_+} \left( a^\dagger - b^\dagger \right)^{n_-} \ket{0}\!\bra{0} \left( c_+^\text{a} \right)^{n_+} \left( c_-^\text{a} \right)^{n_-}.
					\end{split}
				\end{align}

				\subsection{Exceptional point (EP) at $|\gamma_a - \gamma_b| = 4 |g|$}

					When $\gamma_a - \gamma_b = 4 \chi g$, where $\chi = \pm 1$, the Hamiltonian is at an exceptional point (EP), and the Hamiltonian becomes
					\begin{align}
						H = & ~ g \left[- i \left( \chi + \delta \right) a^\dagger a + i \left( \chi - \delta \right) b^\dagger b + a^\dagger b + b^\dagger a\right]\\
						= & ~ g \begin{pmatrix}
							a^\dagger & b^\dagger
						\end{pmatrix} \begin{pmatrix}
							- i \left( \chi + \delta \right) & 1\\
							1 & i \left( \chi - \delta \right)
						\end{pmatrix} \begin{pmatrix}
							a\\
							b
						\end{pmatrix},
					\end{align}
					where $\delta = \Delta \gamma / g$, $\gamma_a = 2 \chi g + 2 \Delta \gamma$, and $\gamma_b = - 2 \chi g + 2 \Delta \gamma$. With a simple recombination of the operators, the Hamiltonian becomes
					\begin{align}
						H = & ~ g \begin{pmatrix}
							d_+^\text{c} & d_-^\text{c}
						\end{pmatrix} \begin{pmatrix}
							i \Delta \gamma & 2 i \\
							0 & i \Delta \gamma
						\end{pmatrix} \begin{pmatrix}
							d_+^\text{a}\\
							d_-^\text{a}
						\end{pmatrix}\\
						= & ~ - i \Delta \gamma \left(d_+^\text{c} d_+^\text{a} + d_-^\text{c} d_-^\text{a}\right) + 2 i g d_+^\text{c} d_-^\text{a},
					\end{align}
					where
					\begin{align}
						d_\pm^\text{c} = \frac{1}{\sqrt{2}} \left( a^\dagger \mp i \chi b^\dagger \right) , \quad d_\pm^\text{a} = \frac{1}{\sqrt{2}} \left( a \pm i \chi b\right).
					\end{align}

					The commutation relations in this case become
					\begin{align}
						\begin{split}
							& \left[ d_\pm^\text{a}, d_\pm^\text{c} \right] = 1,\\
							& \left[ d_\pm^\text{a}, d_\mp^\text{c} \right] = 0,\\
							& \left[ H, d_+^\text{c} \right] = - i \Delta \gamma d_+^\text{c},\\
							& \left[ H, d_-^\text{c} \right] = - i \Delta \gamma d_-^\text{c} + 2 i g d_+^\text{c},\\
							& \left[ H, d_+^\text{a} \right] = i \Delta \gamma d_+^\text{a} - 2 i g d_-^\text{a},\\
							& \left[ H, d_-^\text{a} \right] = i \Delta \gamma d_-^\text{a}.
						\end{split}
					\end{align}

					We can, again, use Eq.~\eqref{Hflat} with $H_\flat = 0$ to find the corresponding vielbein,
					\begin{align}
						\begin{split}
							\mathcal{E} = & \sum_{\substack{n_+ = 0\\ n_- = 0 }}^\infty \frac{h_{n_+ n_-}}{\left(n_+ !\right)^{3 / 2} \left(n_- !\right)^{3 / 2}} \exp \left[\Delta \gamma t \left( n_+ + n_- \right) \right]\\
							& \cdot \left( a^\dagger \right)^{n_-} \left( b^\dagger\right)^{n_+} \ket{0}\!\bra{0} \left( d_+^\text{a} - 2 g t d_-^\text{a}\right)^{n_+} \left( d_-^\text{a}\right)^{n_-},
						\end{split}
					\end{align}
					where the $h_{n_+n_-}$'s are nonzero constants. Therefore, the metric becomes
					\begin{align}
						\begin{split}
							G = & \sum_{\substack{n_+ = 0\\ n_- = 0 }}^\infty \frac{g_{n_+ n_-}}{\left(n_+ !\right)^2 \left(n_- !\right)^2} \exp \left[ 2 \Delta \gamma t \left( n_+ + n_- \right)\right]\\
							& \cdot \left( d_-^{\text{a} \dagger}\right)^{n_-} \left( d_+^{\text{a} \dagger} - 2 g t d_-^{\text{a} \dagger} \right)^{n_+} \ket{0}\\
							& \cdot \bra{0} \left( d_+^\text{a} - 2 g t d_-^\text{a} \right)^{n_+} \left( d_-^\text{a} \right)^{n_-},
						\end{split}
					\end{align}
					where $g_{n_+ n_-} = \left| h_{n_+ n_-} \right|^2$. Once again, we can apply a transformation ${\mathcal{E} \rightarrow \mathcal{E}' = U \mathcal{E}}$, where
					\begin{align}
						\begin{split}
							U = & \sum_{\substack{m = 0\\ n = 0 }}^\infty \frac{\exp \left[itg(m - n) \right]}{\sqrt{2}^{m + n} \left(m !\right) \left(n !\right)}\\
							& \cdot \left( a^\dagger + b^\dagger \right)^{m} \left( a^\dagger - b^\dagger \right)^{n} \ket{0}\!\bra{0} a^{m} b^{n},
						\end{split}
					\end{align}
					so the induced Hamiltonian becomes
					\begin{align}
						H'_\flat = g (a^\dagger b + a b^\dagger).
					\end{align}
					Note that this induced Hamiltonian is indeed Hermitian. This shows that, in addition to the finite dimensional cases, the Hamiltonian of infinite dimension can also be Hermitized via the vielbein formalism.
	\end{appendix}

	\bibliography{References}

\begin{thebibliography}{42}%
\makeatletter
\providecommand \@ifxundefined [1]{%
 \@ifx{#1\undefined}
}%
\providecommand \@ifnum [1]{%
 \ifnum #1\expandafter \@firstoftwo
 \else \expandafter \@secondoftwo
 \fi
}%
\providecommand \@ifx [1]{%
 \ifx #1\expandafter \@firstoftwo
 \else \expandafter \@secondoftwo
 \fi
}%
\providecommand \natexlab [1]{#1}%
\providecommand \enquote  [1]{``#1''}%
\providecommand \bibnamefont  [1]{#1}%
\providecommand \bibfnamefont [1]{#1}%
\providecommand \citenamefont [1]{#1}%
\providecommand \href@noop [0]{\@secondoftwo}%
\providecommand \href [0]{\begingroup \@sanitize@url \@href}%
\providecommand \@href[1]{\@@startlink{#1}\@@href}%
\providecommand \@@href[1]{\endgroup#1\@@endlink}%
\providecommand \@sanitize@url [0]{\catcode `\\12\catcode `\$12\catcode
  `\&12\catcode `\#12\catcode `\^12\catcode `\_12\catcode `\%12\relax}%
\providecommand \@@startlink[1]{}%
\providecommand \@@endlink[0]{}%
\providecommand \url  [0]{\begingroup\@sanitize@url \@url }%
\providecommand \@url [1]{\endgroup\@href {#1}{\urlprefix }}%
\providecommand \urlprefix  [0]{URL }%
\providecommand \Eprint [0]{\href }%
\providecommand \doibase [0]{http://dx.doi.org/}%
\providecommand \selectlanguage [0]{\@gobble}%
\providecommand \bibinfo  [0]{\@secondoftwo}%
\providecommand \bibfield  [0]{\@secondoftwo}%
\providecommand \translation [1]{[#1]}%
\providecommand \BibitemOpen [0]{}%
\providecommand \bibitemStop [0]{}%
\providecommand \bibitemNoStop [0]{.\EOS\space}%
\providecommand \EOS [0]{\spacefactor3000\relax}%
\providecommand \BibitemShut  [1]{\csname bibitem#1\endcsname}%
\let\auto@bib@innerbib\@empty
\bibitem [{\citenamefont {Bender}\ and\ \citenamefont
  {Boettcher}(1998)}]{Bender1998}%
  \BibitemOpen
  \bibinfo {author} {C.~M. Bender}\ and\ \bibinfo {author} {S.~Boettcher},\
  \emph {\bibinfo {title} {Real Spectra in Non-{H}ermitian {H}amiltonians
  Having $\mathcal{PT}$ Symmetry}},\ \href {\doibase
  10.1103/PhysRevLett.80.5243} {\bibfield  {journal} {\bibinfo  {journal}
  {Phys. Rev. Lett.}\ }\textbf {\bibinfo {volume} {80}},\ \bibinfo {pages}
  {5243} (\bibinfo {year} {1998})}\BibitemShut {NoStop}%
\bibitem [{\citenamefont {Bender}\ \emph {et~al.}(2002)\citenamefont {Bender},
  \citenamefont {Brody},\ and\ \citenamefont {Jones}}]{Bender2002}%
  \BibitemOpen
  \bibinfo {author} {C.~M. Bender}, \bibinfo {author} {D.~C. Brody},\ and\
  \bibinfo {author} {H.~F. Jones},\ \emph {\bibinfo {title} {Complex extension
  of quantum mechanics}},\ \href {\doibase 10.1103/PhysRevLett.89.270401}
  {\bibfield  {journal} {\bibinfo  {journal} {Phys. Rev. Lett.}\ }\textbf
  {\bibinfo {volume} {89}},\ \bibinfo {pages} {270401} (\bibinfo {year}
  {2002})}\BibitemShut {NoStop}%
\bibitem [{\citenamefont {Bender}\ \emph {et~al.}(2004)\citenamefont {Bender},
  \citenamefont {Brod}, \citenamefont {Refig},\ and\ \citenamefont
  {Reuter}}]{Bender2004}%
  \BibitemOpen
  \bibinfo {author} {C.~M. Bender}, \bibinfo {author} {J.~Brod}, \bibinfo
  {author} {A.~Refig},\ and\ \bibinfo {author} {M.~E. Reuter},\ \emph {\bibinfo
  {title} {The $\mathcal{C}$ operator in $\mathcal{PT}$-symmetric quantum
  theories}},\ \href {\doibase 10.1088/0305-4470/37/43/009} {\bibfield
  {journal} {\bibinfo  {journal} {J. Phys A: Math. Gen.}\ }\textbf {\bibinfo
  {volume} {37}},\ \bibinfo {pages} {10139} (\bibinfo {year}
  {2004})}\BibitemShut {NoStop}%
\bibitem [{\citenamefont {Bender}(2007)}]{Bender2007}%
  \BibitemOpen
  \bibinfo {author} {C.~M. Bender},\ \emph {\bibinfo {title} {Making sense of
  non-{H}ermitian {H}amiltonians}},\ \href {\doibase
  10.1088/0034-4885/70/6/R03} {\bibfield  {journal} {\bibinfo  {journal} {Rep.
  Prog. Phys.}\ }\textbf {\bibinfo {volume} {70}},\ \bibinfo {pages} {947}
  (\bibinfo {year} {2007})}\BibitemShut {NoStop}%
\bibitem [{\citenamefont {Brody}(2016)}]{Brody2016}%
  \BibitemOpen
  \bibinfo {author} {D.~C. Brody},\ \emph {\bibinfo {title} {Consistency of
  {PT}-symmetric quantum mechanics}},\ \href {\doibase
  10.1088/1751-8113/49/10/10lt03} {\bibfield  {journal} {\bibinfo  {journal}
  {J. Phys. A: Math. Theor.}\ }\textbf {\bibinfo {volume} {49}},\ \bibinfo
  {pages} {10LT03} (\bibinfo {year} {2016})}\BibitemShut {NoStop}%
\bibitem [{\citenamefont {El-Ganainy}\ \emph {et~al.}(2018)\citenamefont
  {El-Ganainy}, \citenamefont {Makris}, \citenamefont {Khajavikhan},
  \citenamefont {Musslimani}, \citenamefont {Rotter},\ and\ \citenamefont
  {Christodoulides}}]{ElGanainy2018}%
  \BibitemOpen
  \bibinfo {author} {R.~El-Ganainy}, \bibinfo {author} {K.~G. Makris}, \bibinfo
  {author} {M.~Khajavikhan}, \bibinfo {author} {Z.~H. Musslimani}, \bibinfo
  {author} {S.~Rotter},\ and\ \bibinfo {author} {D.~N. Christodoulides},\ \emph
  {\bibinfo {title} {Non-Hermitian physics and {PT} symmetry}},\ \href
  {\doibase 10.1038/nphys4323} {\bibfield  {journal} {\bibinfo  {journal} {Nat.
  Phys.}\ }\textbf {\bibinfo {volume} {14}},\ \bibinfo {pages} {11} (\bibinfo
  {year} {2018})}\BibitemShut {NoStop}%
\bibitem [{\citenamefont {Bagarello}\ \emph {et~al.}(2016)\citenamefont
  {Bagarello}, \citenamefont {Passante},\ and\ \citenamefont
  {Trapani}}]{Bagarello2016}%
  \BibitemOpen
  \bibinfo {editor} {F.~Bagarello}, \bibinfo {editor} {R.~Passante},\ and\
  \bibinfo {editor} {C.~Trapani},\ eds.,\ \href {\doibase
  10.1007/978-3-319-31356-6} {\emph {\bibinfo {title} {Non-Hermitian
  Hamiltonians in Quantum Physics}}}\ (\bibinfo  {publisher} {Springer
  International Publishing},\ \bibinfo {year} {2016})\BibitemShut {NoStop}%
\bibitem [{\citenamefont {Ashida}\ \emph {et~al.}(2020)\citenamefont {Ashida},
  \citenamefont {Gong},\ and\ \citenamefont {Ueda}}]{Ashida2020}%
  \BibitemOpen
  \bibinfo {author} {Y.~Ashida}, \bibinfo {author} {Z.~Gong},\ and\ \bibinfo
  {author} {M.~Ueda},\ \emph {\bibinfo {title} {Non-Hermitian physics}},\ \href
  {\doibase 10.1080/00018732.2021.1876991} {\bibfield  {journal} {\bibinfo
  {journal} {Adv. Phys.}\ }\textbf {\bibinfo {volume} {69}},\ \bibinfo {pages}
  {249} (\bibinfo {year} {2020})}\BibitemShut {NoStop}%
\bibitem [{\citenamefont {Tzeng}\ \emph {et~al.}(2021)\citenamefont {Tzeng},
  \citenamefont {Ju}, \citenamefont {Chen},\ and\ \citenamefont
  {Huang}}]{Tzeng2021}%
  \BibitemOpen
  \bibinfo {author} {Y.-C. Tzeng}, \bibinfo {author} {C.-Y. Ju}, \bibinfo
  {author} {G.-Y. Chen},\ and\ \bibinfo {author} {W.-M. Huang},\ \emph
  {\bibinfo {title} {Hunting for the non-Hermitian exceptional points with
  fidelity susceptibility}},\ \href {\doibase 10.1103/PhysRevResearch.3.013015}
  {\bibfield  {journal} {\bibinfo  {journal} {Phys. Rev. Res.}\ }\textbf
  {\bibinfo {volume} {3}},\ \bibinfo {pages} {013015} (\bibinfo {year}
  {2021})}\BibitemShut {NoStop}%
\bibitem [{\citenamefont {Tu}\ \emph {et~al.}()\citenamefont {Tu},
  \citenamefont {Jang}, \citenamefont {Chang},\ and\ \citenamefont
  {Tzeng}}]{Tu2022}%
  \BibitemOpen
  \bibinfo {author} {Y.-T. Tu}, \bibinfo {author} {I.~Jang}, \bibinfo {author}
  {P.-Y. Chang},\ and\ \bibinfo {author} {Y.-C. Tzeng},\ \emph {\bibinfo
  {title} {General properties of fidelity in non-Hermitian quantum systems with
  PT symmetry}},\ \href@noop {} {\ }\Eprint {http://arxiv.org/abs/2203.01834v1}
  {2203.01834v1} \BibitemShut {NoStop}%
\bibitem [{\citenamefont {Mostafazadeh}(2003)}]{Mostafazadeh2003}%
  \BibitemOpen
  \bibinfo {author} {A.~Mostafazadeh},\ \emph {\bibinfo {title}
  {Pseudo-{H}ermiticity and generalized $\mathcal{PT}$- and
  $\mathcal{CPT}$-symmetries}},\ \href {\doibase 10.1063/1.1539304} {\bibfield
  {journal} {\bibinfo  {journal} {J. Math. Phys.}\ }\textbf {\bibinfo {volume}
  {44}},\ \bibinfo {pages} {974} (\bibinfo {year} {2003})}\BibitemShut
  {NoStop}%
\bibitem [{\citenamefont {Mostafazadeh}(2004)}]{Mostafazadeh2004}%
  \BibitemOpen
  \bibinfo {author} {A.~Mostafazadeh},\ \emph {\bibinfo {title} {{Time
  dependent Hilbert spaces, geometric phases, and general covariance in quantum
  mechanics}}},\ \href {\doibase 10.1016/j.physleta.2003.12.008} {\bibfield
  {journal} {\bibinfo  {journal} {Phys. Lett. A}\ }\textbf {\bibinfo {volume}
  {320}},\ \bibinfo {pages} {375} (\bibinfo {year} {2004})}\BibitemShut
  {NoStop}%
\bibitem [{\citenamefont {Brody}(2013)}]{Brody2013}%
  \BibitemOpen
  \bibinfo {author} {D.~C. Brody},\ \emph {\bibinfo {title} {Biorthogonal
  quantum mechanics}},\ \href {\doibase 10.1088/1751-8113/47/3/035305}
  {\bibfield  {journal} {\bibinfo  {journal} {J. Phys. A: Math. Theor.}\
  }\textbf {\bibinfo {volume} {47}},\ \bibinfo {pages} {035305} (\bibinfo
  {year} {2013})}\BibitemShut {NoStop}%
\bibitem [{\citenamefont {Znojil}(2016)}]{Znojil2016}%
  \BibitemOpen
  \bibinfo {author} {M.~Znojil},\ \emph {\bibinfo {title} {{Is
  $\mathcal{PT}$-symmetric quantum theory false as a fundamental theory?}}},\
  \href {\doibase 10.14311/AP.2016.56.0254} {\bibfield  {journal} {\bibinfo
  {journal} {Acta Polytechnica (Prague)}\ }\textbf {\bibinfo {volume} {56}},\
  \bibinfo {pages} {254} (\bibinfo {year} {2016})}\BibitemShut {NoStop}%
\bibitem [{\citenamefont {Ju}\ \emph {et~al.}(2019)\citenamefont {Ju},
  \citenamefont {Miranowicz}, \citenamefont {Chen},\ and\ \citenamefont
  {Nori}}]{Ju2019}%
  \BibitemOpen
  \bibinfo {author} {C.-Y. Ju}, \bibinfo {author} {A.~Miranowicz}, \bibinfo
  {author} {G.-Y. Chen},\ and\ \bibinfo {author} {F.~Nori},\ \emph {\bibinfo
  {title} {Non-{H}ermitian {H}amiltonians and no-go theorems in quantum
  information}},\ \href {\doibase 10.1103/physreva.100.062118} {\bibfield
  {journal} {\bibinfo  {journal} {Phys. Rev. A}\ }\textbf {\bibinfo {volume}
  {100}},\ \bibinfo {pages} {062118} (\bibinfo {year} {2019})}\BibitemShut
  {NoStop}%
\bibitem [{\citenamefont {Einstein}(2014)}]{Einstein2014}%
  \BibitemOpen
  \bibinfo {author} {A.~Einstein},\ \href@noop {} {\emph {\bibinfo {title} {The
  Meaning of Relativity: Including the Relativistic Theory of the Non-Symmetric
  Field}}},\ \bibinfo {edition} {5th}\ ed.\ (\bibinfo  {publisher} {Princeton
  Univers. Press},\ \bibinfo {year} {2014})\BibitemShut {NoStop}%
\bibitem [{\citenamefont {Chan}\ and\ \citenamefont {Chua}(2000)}]{Chan2000}%
  \BibitemOpen
  \bibinfo {author} {C.-T. Chan}\ and\ \bibinfo {author} {C.-K. Chua},\ \emph
  {\bibinfo {title} {Equivalent Relations between Quantum Dynamics as Derived
  from a Gauge Transformation}},\ \href@noop {} {\  (\bibinfo {year} {2000})},\
  \Eprint {http://arxiv.org/abs/hep-th/0009040} {arXiv:hep-th/0009040 [hep-th]}
  \BibitemShut {NoStop}%
\bibitem [{\citenamefont {Nakahara}(2003)}]{Nakahara2003}%
  \BibitemOpen
  \bibinfo {author} {M.~Nakahara},\ \href@noop {} {\emph {\bibinfo {title}
  {Geometry, Topology and Physics}}},\ \bibinfo {edition} {2nd}\ ed.\ (\bibinfo
   {publisher} {{IOP Publishing, Bristol}},\ \bibinfo {year}
  {2003})\BibitemShut {NoStop}%
\bibitem [{\citenamefont {Misner}\ \emph {et~al.}(2017)\citenamefont {Misner},
  \citenamefont {Thorne},\ and\ \citenamefont {Wheeler}}]{Misner2017}%
  \BibitemOpen
  \bibinfo {author} {C.~W. Misner}, \bibinfo {author} {K.~S. Thorne},\ and\
  \bibinfo {author} {J.~A. Wheeler},\ \href
  {https://www.ebook.de/de/product/28672241/charles_w_misner_kip_s_thorne_john_archibald_wheeler_gravitation.html}
  {\emph {\bibinfo {title} {Gravitation}}}\ (\bibinfo  {publisher} {Princeton
  Univers. Press},\ \bibinfo {year} {2017})\BibitemShut {NoStop}%
\bibitem [{\citenamefont {Wald}(1984)}]{Wald1984}%
  \BibitemOpen
  \bibinfo {author} {R.~M. Wald},\ \href
  {https://www.ebook.de/de/product/3238000/robert_m_wald_general_relativity.html}
  {\emph {\bibinfo {title} {General Relativity}}}\ (\bibinfo  {publisher}
  {University of Chicago Press},\ \bibinfo {year} {1984})\BibitemShut {NoStop}%
\bibitem [{\citenamefont {van Nieuwenhuizen}(1981)}]{Nieuwenhuizen1981}%
  \BibitemOpen
  \bibinfo {author} {P.~van Nieuwenhuizen},\ \emph {\bibinfo {title}
  {Supergravity}},\ \href {\doibase 10.1016/0370-1573(81)90157-5} {\bibfield
  {journal} {\bibinfo  {journal} {Phys. Rep.}\ }\textbf {\bibinfo {volume}
  {68}},\ \bibinfo {pages} {189} (\bibinfo {year} {1981})}\BibitemShut
  {NoStop}%
\bibitem [{\citenamefont {West}(1990)}]{West1990}%
  \BibitemOpen
  \bibinfo {author} {P.~West},\ \href {\doibase 10.1142/1002} {\emph {\bibinfo
  {title} {Introduction to Supersymmetry and Supergravity}}}\ (\bibinfo
  {publisher} {World Scientific},\ \bibinfo {year} {1990})\BibitemShut
  {NoStop}%
\bibitem [{\citenamefont {Siegel}(1993{\natexlab{a}})}]{Siegel1993}%
  \BibitemOpen
  \bibinfo {author} {W.~Siegel},\ \emph {\bibinfo {title} {Two-vierbein
  formalism for string-inspired axionic gravity}},\ \href {\doibase
  10.1103/physrevd.47.5453} {\bibfield  {journal} {\bibinfo  {journal} {Phys.
  Rev. D}\ }\textbf {\bibinfo {volume} {47}},\ \bibinfo {pages} {5453}
  (\bibinfo {year} {1993}{\natexlab{a}})}\BibitemShut {NoStop}%
\bibitem [{\citenamefont {Siegel}(1993{\natexlab{b}})}]{Siegel1993a}%
  \BibitemOpen
  \bibinfo {author} {W.~Siegel},\ \emph {\bibinfo {title} {Superspace duality
  in low-energy superstrings}},\ \href {\doibase 10.1103/physrevd.48.2826}
  {\bibfield  {journal} {\bibinfo  {journal} {Phys. Rev. D}\ }\textbf {\bibinfo
  {volume} {48}},\ \bibinfo {pages} {2826} (\bibinfo {year}
  {1993}{\natexlab{b}})}\BibitemShut {NoStop}%
\bibitem [{\citenamefont {Pol{\'{a}}{\v{c}}ek}\ and\ \citenamefont
  {Siegel}(2014)}]{Polacek2014}%
  \BibitemOpen
  \bibinfo {author} {M.~Pol{\'{a}}{\v{c}}ek}\ and\ \bibinfo {author}
  {W.~Siegel},\ \emph {\bibinfo {title} {T-duality off shell in 3D type {II}
  superspace}},\ \href {\doibase 10.1007/jhep06(2014)107} {\bibfield  {journal}
  {\bibinfo  {journal} {J. High Energy Phys.}\ }\textbf {\bibinfo {volume}
  {2014}},\ \bibinfo {pages} {107} (\bibinfo {year} {2014})}\BibitemShut
  {NoStop}%
\bibitem [{\citenamefont {Ju}\ and\ \citenamefont {Siegel}(2016)}]{Ju2016}%
  \BibitemOpen
  \bibinfo {author} {C.-Y. Ju}\ and\ \bibinfo {author} {W.~Siegel},\ \emph
  {\bibinfo {title} {Gauging unbroken symmetries in {F}-theory}},\ \href
  {\doibase 10.1103/physrevd.94.106004} {\bibfield  {journal} {\bibinfo
  {journal} {Phys. Rev. D}\ }\textbf {\bibinfo {volume} {94}},\ \bibinfo
  {pages} {106004} (\bibinfo {year} {2016})}\BibitemShut {NoStop}%
\bibitem [{\citenamefont {Linch}\ and\ \citenamefont
  {Siegel}(2021)}]{Linch2021}%
  \BibitemOpen
  \bibinfo {author} {W.~D. Linch}\ and\ \bibinfo {author} {W.~Siegel},\ \emph
  {\bibinfo {title} {F-theory from fundamental five-branes}},\ \href {\doibase
  10.1007/jhep02(2021)047} {\bibfield  {journal} {\bibinfo  {journal} {J. High
  Energy Phys.}\ }\textbf {\bibinfo {volume} {2021}},\ \bibinfo {pages} {47}
  (\bibinfo {year} {2021})}\BibitemShut {NoStop}%
\bibitem [{\citenamefont {Znojil}(2008)}]{Znojil2008}%
  \BibitemOpen
  \bibinfo {author} {M.~Znojil},\ \emph {\bibinfo {title} {Time-dependent
  version of crypto-{H}ermitian quantum theory}},\ \href {\doibase
  10.1103/physrevd.78.085003} {\bibfield  {journal} {\bibinfo  {journal} {Phys.
  Rev. D}\ }\textbf {\bibinfo {volume} {78}},\ \bibinfo {pages} {085003}
  (\bibinfo {year} {2008})}\BibitemShut {NoStop}%
\bibitem [{\citenamefont {Mostahazadeh}(2010)}]{Mostafazadeh2010}%
  \BibitemOpen
  \bibinfo {author} {A.~Mostahazadeh},\ \emph {\bibinfo {title}
  {Pseudo-{H}ermitian Reresentation of Quantum Mechanics}},\ \href {\doibase
  10.1142/s0219887810004816} {\bibfield  {journal} {\bibinfo  {journal} {Int.
  J. Geom. Methods Mod. Phys.}\ }\textbf {\bibinfo {volume} {07}},\ \bibinfo
  {pages} {1191} (\bibinfo {year} {2010})}\BibitemShut {NoStop}%
\bibitem [{\citenamefont {Hamazaki}\ \emph {et~al.}(2020)\citenamefont
  {Hamazaki}, \citenamefont {Kawabata}, \citenamefont {Kura},\ and\
  \citenamefont {Ueda}}]{Hamazaki2020}%
  \BibitemOpen
  \bibinfo {author} {R.~Hamazaki}, \bibinfo {author} {K.~Kawabata}, \bibinfo
  {author} {N.~Kura},\ and\ \bibinfo {author} {M.~Ueda},\ \emph {\bibinfo
  {title} {Universality classes of non-Hermitian random matrices}},\ \href
  {\doibase 10.1103/physrevresearch.2.023286} {\bibfield  {journal} {\bibinfo
  {journal} {Phys. Rev. Res.}\ }\textbf {\bibinfo {volume} {2}},\ \bibinfo
  {pages} {023286} (\bibinfo {year} {2020})}\BibitemShut {NoStop}%
\bibitem [{\citenamefont {Ohlsson}\ and\ \citenamefont
  {Zhou}(2021)}]{Ohlsson2021}%
  \BibitemOpen
  \bibinfo {author} {T.~Ohlsson}\ and\ \bibinfo {author} {S.~Zhou},\ \emph
  {\bibinfo {title} {Density-matrix formalism for $\cal{PT}$-symmetric
  non-{H}ermitian {H}amiltonians with the {L}indblad equation}},\ \href
  {\doibase 10.1103/physreva.103.022218} {\bibfield  {journal} {\bibinfo
  {journal} {Phys. Rev. A}\ }\textbf {\bibinfo {volume} {103}},\ \bibinfo
  {pages} {022218} (\bibinfo {year} {2021})}\BibitemShut {NoStop}%
\bibitem [{\citenamefont {Lee}\ \emph {et~al.}(2014)\citenamefont {Lee},
  \citenamefont {Hsieh}, \citenamefont {Flammia},\ and\ \citenamefont
  {Lee}}]{Lee2014}%
  \BibitemOpen
  \bibinfo {author} {Y.-C. Lee}, \bibinfo {author} {M.-H. Hsieh}, \bibinfo
  {author} {S.~T. Flammia},\ and\ \bibinfo {author} {R.-K. Lee},\ \emph
  {\bibinfo {title} {Local $\mathcal{PT}$ Symmetry Violates the No-Signaling
  Principle}},\ \href {\doibase 10.1103/PhysRevLett.112.130404} {\bibfield
  {journal} {\bibinfo  {journal} {Phys. Rev. Lett.}\ }\textbf {\bibinfo
  {volume} {112}},\ \bibinfo {pages} {130404} (\bibinfo {year}
  {2014})}\BibitemShut {NoStop}%
\bibitem [{\citenamefont {Pati}()}]{Pati2014}%
  \BibitemOpen
  \bibinfo {author} {A.~K. Pati},\ \emph {\bibinfo {title} {Violation of
  Invariance of Entanglement Under Local PT Symmetric Unitary}},\ \href
  {http://arxiv.org/abs/1404.6166} {\ }\Eprint
  {http://arxiv.org/abs/arXiv:1404.6166} {arXiv:1404.6166} \BibitemShut
  {NoStop}%
\bibitem [{\citenamefont {Chen}\ \emph {et~al.}(2014)\citenamefont {Chen},
  \citenamefont {Chen},\ and\ \citenamefont {Chen}}]{Chen2014}%
  \BibitemOpen
  \bibinfo {author} {S.-L. Chen}, \bibinfo {author} {G.-Y. Chen},\ and\
  \bibinfo {author} {Y.-N. Chen},\ \emph {\bibinfo {title} {Increase of
  entanglement by local $\mathcal{PT}$-symmetric operations}},\ \href {\doibase
  10.1103/PhysRevA.90.054301} {\bibfield  {journal} {\bibinfo  {journal} {Phys.
  Rev. A}\ }\textbf {\bibinfo {volume} {90}},\ \bibinfo {pages} {054301}
  (\bibinfo {year} {2014})}\BibitemShut {NoStop}%
\bibitem [{\citenamefont {Stefano}\ \emph {et~al.}(2019)\citenamefont
  {Stefano}, \citenamefont {Settineri}, \citenamefont {Macr{\`{\i}}},
  \citenamefont {Garziano}, \citenamefont {Stassi}, \citenamefont {Savasta},\
  and\ \citenamefont {Nori}}]{Stefano2019}%
  \BibitemOpen
  \bibinfo {author} {O.~D. Stefano}, \bibinfo {author} {A.~Settineri}, \bibinfo
  {author} {V.~Macr{\`{\i}}}, \bibinfo {author} {L.~Garziano}, \bibinfo
  {author} {R.~Stassi}, \bibinfo {author} {S.~Savasta},\ and\ \bibinfo {author}
  {F.~Nori},\ \emph {\bibinfo {title} {Resolution of gauge ambiguities in
  ultrastrong-coupling cavity quantum electrodynamics}},\ \href {\doibase
  10.1038/s41567-019-0534-4} {\bibfield  {journal} {\bibinfo  {journal} {Nat.
  Phys.}\ }\textbf {\bibinfo {volume} {15}},\ \bibinfo {pages} {803} (\bibinfo
  {year} {2019})}\BibitemShut {NoStop}%
\bibitem [{\citenamefont {Garziano}\ \emph {et~al.}(2020)\citenamefont
  {Garziano}, \citenamefont {Settineri}, \citenamefont {Stefano}, \citenamefont
  {Savasta},\ and\ \citenamefont {Nori}}]{Garziano2020}%
  \BibitemOpen
  \bibinfo {author} {L.~Garziano}, \bibinfo {author} {A.~Settineri}, \bibinfo
  {author} {O.~D. Stefano}, \bibinfo {author} {S.~Savasta},\ and\ \bibinfo
  {author} {F.~Nori},\ \emph {\bibinfo {title} {Gauge invariance of the Dicke
  and Hopfield models}},\ \href {\doibase 10.1103/physreva.102.023718}
  {\bibfield  {journal} {\bibinfo  {journal} {Phys. Rev. A}\ }\textbf {\bibinfo
  {volume} {102}},\ \bibinfo {pages} {023718} (\bibinfo {year}
  {2020})}\BibitemShut {NoStop}%
\bibitem [{\citenamefont {Settineri}\ \emph {et~al.}(2021)\citenamefont
  {Settineri}, \citenamefont {Stefano}, \citenamefont {Zueco}, \citenamefont
  {Hughes}, \citenamefont {Savasta},\ and\ \citenamefont
  {Nori}}]{Settineri2021}%
  \BibitemOpen
  \bibinfo {author} {A.~Settineri}, \bibinfo {author} {O.~D. Stefano}, \bibinfo
  {author} {D.~Zueco}, \bibinfo {author} {S.~Hughes}, \bibinfo {author}
  {S.~Savasta},\ and\ \bibinfo {author} {F.~Nori},\ \emph {\bibinfo {title}
  {Gauge freedom, quantum measurements, and time-dependent interactions in
  cavity {QED}}},\ \href {\doibase 10.1103/physrevresearch.3.023079} {\bibfield
   {journal} {\bibinfo  {journal} {Phys. Rev. Res.}\ }\textbf {\bibinfo
  {volume} {3}},\ \bibinfo {pages} {023079} (\bibinfo {year}
  {2021})}\BibitemShut {NoStop}%
\bibitem [{\citenamefont {Savasta}\ \emph {et~al.}(2021)\citenamefont
  {Savasta}, \citenamefont {Stefano}, \citenamefont {Settineri}, \citenamefont
  {Zueco}, \citenamefont {Hughes},\ and\ \citenamefont {Nori}}]{Savasta2021}%
  \BibitemOpen
  \bibinfo {author} {S.~Savasta}, \bibinfo {author} {O.~D. Stefano}, \bibinfo
  {author} {A.~Settineri}, \bibinfo {author} {D.~Zueco}, \bibinfo {author}
  {S.~Hughes},\ and\ \bibinfo {author} {F.~Nori},\ \emph {\bibinfo {title}
  {Gauge principle and gauge invariance in two-level systems}},\ \href
  {\doibase 10.1103/physreva.103.053703} {\bibfield  {journal} {\bibinfo
  {journal} {Phys. Rev. A}\ }\textbf {\bibinfo {volume} {103}},\ \bibinfo
  {pages} {053703} (\bibinfo {year} {2021})}\BibitemShut {NoStop}%
\bibitem [{\citenamefont {Kato}(1976)}]{Kato1976}%
  \BibitemOpen
  \bibinfo {author} {T.~Kato},\ \href {https://cds.cern.ch/record/101545}
  {\emph {\bibinfo {title} {Perturbation theory for linear operators}}},\
  \bibinfo {edition} {2nd}\ ed.,\ Grundlehren der mathematischen
  Wissenschaften: a series of comprehensive studies in mathematics\ (\bibinfo
  {publisher} {Springer},\ \bibinfo {address} {Berlin},\ \bibinfo {year}
  {1976})\BibitemShut {NoStop}%
\bibitem [{\citenamefont {Heiss}(2004)}]{Heiss2004}%
  \BibitemOpen
  \bibinfo {author} {W.~D. Heiss},\ \emph {\bibinfo {title} {Exceptional points
  of non-{H}ermitian operators}},\ \href {\doibase 10.1088/0305-4470/37/6/034}
  {\bibfield  {journal} {\bibinfo  {journal} {J. Phys A: Math. Gen.}\ }\textbf
  {\bibinfo {volume} {37}},\ \bibinfo {pages} {2455} (\bibinfo {year}
  {2004})}\BibitemShut {NoStop}%
\bibitem [{\citenamefont {\"{O}zdemir}\ \emph {et~al.}(2019)\citenamefont
  {\"{O}zdemir}, \citenamefont {Rotter}, \citenamefont {Nori},\ and\
  \citenamefont {Yang}}]{Oezdemir2019}%
  \BibitemOpen
  \bibinfo {author} {{\c{S}}.~K. \"{O}zdemir}, \bibinfo {author} {S.~Rotter},
  \bibinfo {author} {F.~Nori},\ and\ \bibinfo {author} {L.~Yang},\ \emph
  {\bibinfo {title} {Parity{\textendash}time symmetry and exceptional points in
  photonics}},\ \href {\doibase 10.1038/s41563-019-0304-9} {\bibfield
  {journal} {\bibinfo  {journal} {Nat. Mater.}\ }\textbf {\bibinfo {volume}
  {18}},\ \bibinfo {pages} {783} (\bibinfo {year} {2019})}\BibitemShut
  {NoStop}%
\bibitem [{\citenamefont {Minganti}\ \emph {et~al.}(2019)\citenamefont
  {Minganti}, \citenamefont {Miranowicz}, \citenamefont {Chhajlany},\ and\
  \citenamefont {Nori}}]{Minganti2019}%
  \BibitemOpen
  \bibinfo {author} {F.~Minganti}, \bibinfo {author} {A.~Miranowicz}, \bibinfo
  {author} {R.~W. Chhajlany},\ and\ \bibinfo {author} {F.~Nori},\ \emph
  {\bibinfo {title} {Quantum exceptional points of non-{H}ermitian
  {H}amiltonians and {L}iouvillians: The effects of quantum jumps}},\ \href
  {\doibase 10.1103/physreva.100.062131} {\bibfield  {journal} {\bibinfo
  {journal} {Phys. Rev. A}\ }\textbf {\bibinfo {volume} {100}},\ \bibinfo
  {pages} {062131} (\bibinfo {year} {2019})}\BibitemShut {NoStop}%
\end{thebibliography}%

\end{document}